\documentstyle[12pt]{article}

\begin{document}
\newcommand {\be}{\begin{equation}}
\newcommand {\alp}{\alpha'}
\newcommand {\Dl}{\Delta}
\newcommand {\ee}{\end{equation}}
\newcommand {\bea}{\begin{array}}
\newcommand {\cl}{\centerline}
\newcommand {\eea}{\end{array}}
\baselineskip 0.65 cm
\begin{flushright}
IPM-97-\\
hep-th/9704007
\end{flushright}
\begin{center}
\bf \Large {D-brane Thickness}
\end{center}
\begin{center}
H. Arfaei\footnote{e-mail: Arfaei@ theory.ipm.ac.ir} 
$\;\;\;\;\;$ and $\;\;\;\;$
M.M. Sheikh Jabbari\footnote{e-mail: jabbari@netware2.ipm.ac.ir}
\end{center}
\cl {\it Institute For Studies in Theoretical Physics and Mathematics 
IPM} \cl{\it P.O.Box 19395-5531, Tehran,Iran}
\cl{\it Department of Physics Sharif University of Technology}
\cl{\it P.O.Box 11365-9161, Tehran, Iran}
\vskip 2cm
\begin{abstract}
 The scale in the D-brane theory is discussed from different points
of view. It is shown that scattering of a $D_4$-brane by a $D_0$-brane
gives the condition to penetrate beyond the string scale.

\end{abstract}

\newpage
 By now D-branes are accepted to represent extended sources of different $RR$
charges satisfying the $BPS$ conditions [1,2]. These objects necessarily 
exist and are needed to explain certain duality relations. Some properties of 
these objects can be studied within the perturbative string theory but for 
their detailed finer properties we must certainly wait for a more fundamental 
or exact formulation such as M-theory or truely second quantized string theory. 
The point that these objects are more stringy or fundamental description of low 
energy limit of extended objects in the SUGRA (p-branes), are ad ocated by many 
authors [3,4]. On the other hand there are claims that D-branes being $RR$ 
solitons, have masses proportional to $1/g$ and hence length scales which are 
smaller by a factor of $g$. If this is true D-branes can be used to probe 
scales smaller than string scale and hence provide some means to go beyond the 
bounds set by stringy uncertainty principle.

 This principle is based on the point that due to natural string length the
uncertainty in the size of the string used as probe is bounded. If we use the
high energy states natural to probe shorter distances we have to include string
states of larger size [5,6]. In this short note we argue that in most cases 
despite their mass being larger by a factor of $g$,
D-branes show the same scales as ordinary string states.
Due to the capacity of emitting massive states, they form a stringy halo 
with a thickness of the same order as string scale. When $g$ is large such that
the compton wave length is larger than the string scale the new scale which is
no longer, shorter, becomes important.
 We shall consider the problem from three different points of view all related 
to the short distance structure. 
 First we show if a brane is probed by another brane , of the same or different 
dimension, the effective potential or amplitude shows a charactristic length of 
the same order of the string scale, except when a $D_0$-brane probes a 
$D_0$-brane or a $D_4$-brane. As we will discuss here 
in these special cases of D-brane scattering, the stringy halo around 
a D-brane is not detected (due to SUSY), hence we can probe beyond string scale 
in these cases. Moreover in these cases scattering amplitude vanishes to all
orders of closed string spectrum exchanged between branes, which is a sign of 
SUSY in open string channel.

 In the study of space-time behaviour of D-branes from string theoretical 
point of view attention has been focused largely on large distance behaviour of 
the brane 
interactions where the results of low energy supergravity is regained. 
To reveal the stringy nature of the D-branes one should consider their short 
distance properties which may have surprises. 
This scale shows itself in the 
divergence of the amplitude of the closed string exchange between branes. 
 Secondly we consider the effective potential of a single brane due to 
emission of a quanta of massive or massless strings. 
 The same mechanism that leads to the Hagedron temprature is responsible for 
the divergence of the effective potential at a critical distance from 
the brane. 
The phenomena is then clearly rooted in the growth of degeneracy of states
with increasing mass. In the third approach, we consider open strings with the 
ends on 
the same brane; they also provide a halo with the string scale. This scale is 
probed by scattering of zero mass strings off the branes as was considered by
Klebanov etal.[7,8]. 
In the following we discuss the above points of view in more
detail. The discussion is in the context of type II theories.
\\

{\bf i)D-brane short distance interactions:}

 As it has been discussed in [1,2] D-brane, D-brane interaction in string 
theoretical description is obtained by one loop vacuum amplitude of open 
strings stretched between branes
\be
 A=\int {dt \over t} Tr e^{-tH}
\ee
where $H$ is the open string Hamiltonian:
\be
H=P^2 + M^2, \;\;\;\;\;\;\ \alpha'M^2=N+a+{ Y^2 \over 4\pi^2 \alpha'}
\ee
where $N$ is the excitation number and $a$ the ordering constant and $Y$ their  
relative distance.
 In order to calculate the short distance behaviour of the amplitude we must
study large $t$ limit of the integrand, so we consider different relative branes 
and in each case discuss the short distance behaviour:

 {\bf A)Parallel D-branes}:
 In this case $a=0$ for R sector and $a=-{1\over2}$ for NS sector and as it is 
shown in [2] amplitude vanishes for all orders of $t$.
This means that the cancellation occurs in any relative distance for similar,
parallel D-branes, which is a sign of supersymmetry of open string spectrum.
So let us consider $NSNS$ sector contribution (all the discussion given here 
holds for $RR$ sector too.). Although in this case the different contributions
cancel  but as we will see slight perturbation in brane's relative orientation
make the amplitude divergent, and in the case of anti-parallel branes the 
dominant part at short distance comes from the   
large $t$ part of the integrand which can be separated as [9]:  

\be
  A=2V_{p+1}{1 \over 2} \int_T^{\infty} {dt \over t}(8\pi^2\alpha' t)^{{-(p+1) \over 2}} 
  e^{-t(\frac {Y^2-2\pi^2\alpha'}{2\pi \alpha'})}
\ee
where $T$ is chosen so that $e^{\pi T}>>T^4$.

 As we see amplitude becomes divergent for $Y^2< 2\pi^2 \alp$, i.e. T can be 
choosen roughly larger than 10.

 The amplitude diverges for
\be
 Y^2 < Y_c^2=\sqrt{2\pi^2 \alp}\sim l_s. 
\ee
 This divergence could be explained both in terms of closed strings exchanged
between branes and also open strings going around a loop. 
 In open string channel at $Y=Y_c$ open strings become massless and beyond this 
limit they become tachyonic, but in the case of parallel D-branes the 
cancellation  removes their contribution and restores SUSY (The configuration 
of two similar D-branes preserves half of the 32 super charges already present 
in type-II theories).      
In closed strings channel  at this limit massive modes 
dominate (massive closed strings are dual to massless or tachyonic open strings 
[5]).
This point could be understood  roughly as follows:
D-brane interacts via exchange of all the closed string spectrum, containing
massive modes too. The propagator of these strings decays exponentially:
$exp(-m|Y|)G_{9-p}(Y)$, but the degeneracy of states grows as $ ({m \over m_0})^{-11/4}exp{m \over m_0}$.
Hence in closed string channel amplitude is written roughly as:
\be
  A= \sum_{spect.}(V_{p+1}T_p^2) \rho (m) e^{m|Y|}G_{9-p}(Y)
\ee
where $T_p$ is the D-brane tension. For large $m$ terms, this sum behaves as
\be
 A \sim T_p^2 V_{p+1} \sum ({m \over m_0})^{-11/4} e^{-m(|Y| - {1 \over m_0})}.
\ee

which is divergent for 
\be 
|Y|<{1 \over m_0}=Y_c=T_H^{-1}
\ee
where $T_H$ is the Hagedron temprature. 
\\

{\bf B)Parallel $D_p,D_{p'}$-branes:}
It is shown that the interaction amplitude of p and p' branes is given 
by [9,10]:

\be 
A= V_{p'+1} \int {dt \over t}(8\pi^2\alp t)^{-({p'+1 \over 2})}
e^{-t{Y^2 \over 2\pi\alp}}(NS-R).
\ee
where
\be
NS-R= {1\over 2}(2\pi)^{4-\Delta/2}\biggl( (\frac{\theta_3}{\theta'_1})^{4-\Delta/2}
(\frac{\theta_2}{\theta_4})^{\Delta/2}-(\frac{\theta_2}{\theta'_1})^{4-\Delta/2}
(\frac{\theta_3}{\theta_4})^{\Delta/2}\biggr)\;\;\;\;;\;\;\;\  \Delta=p-p'.
\ee
 
 The short distance behaviour of this amplitude is dominated by large t. 
At large $t$ limit integral can be estimated as: 
\be
A \sim V_{p'+1} \int_T^{\infty} {dt \over 2t}(8\pi^2\alp t)^{-({p'+1 \over 2})}
e^{-t\bigl({Y^2 \over 2\pi\alp}-(1-\Delta/4)\pi\bigr)}. 
\ee
 showing a divergence at
\be
Y_c^2=2\pi^2\alp(1-\Delta/4).
\ee
The amplitude converges for $|Y| \geq Y_c$.

This $Y_c$ shows that we can not probe a D-brane beyond the string scale with
a different kind of D-brane; D-brane shows a thickness of the same order of, 
string scale. 
 This is expected if D-branes are thought to be string solitonic 
solutions. 
In the open string channel $Y_c$ is the length where open string 
concept fails and the tachyons return to the theory. 
Potential beyond this critical length can't be studied by analytic extension  
as pointed in [9]. 
We shall consider this issuse in detail later.
 
 In the special case of $\Delta=4$, $Y_c=0$, and the amplitude vanishes for all 
orders of $t$, and in this case again $N=1$ SUSY of open strings is restored [12], 
so stringy halo around a brane becomes invisible to the other brane.
Again as in the case of similar branes any small change in the relative 
orientation brings the divergence back but the small interaction observed 
when in non-relativistic motion also does not show any non-zero critical 
distance.
\\

{\bf C)Branes at angle:}
 In the case of two similar branes at angle $\pi\theta$ [9], the critical 
distance $Y_c$ is non-zero and is equal to $Y_c^2={l_s^2 \theta \over \sqrt2}$. 
This shows that in actual probing experiment when one has to deal with a wave 
packet having contribution of different angles we again encountor the string
scale except for $D_0$-brane case where relative angle has no meaning.

 In the case of p,p' branes at relative angle $\pi\theta$ [9], $Y_c$ is obtained
to be
\be 
Y_c^2=({1\over 2}-{\Delta \over 4}+|\theta-1/2|)l_s^2.
\ee
 In general such two branes can probe each other only up to string scale but for 
the case of $(\Dl=2,\theta=1/2)$ and $(\Dl=4,\theta=0)$ $Y_c$ vanishes. In
this case wave packet of the smaller brane again having contributions from 
different angles , blocks the probing beyond string scales. Although this  
argument does not work when a zero brane probes a four-brane: i.e. string scale 
is supressed in this case.
\\
{\bf ii) Potential near $Y_c$:} 
 
 Open string concept loses its efficiency beyond $Y_c$ where they 
become tachyonic, the phenomena responsible for the Hagedron temprature makes 
the amplitude divergent for all values $Y<Y_c$, blocking the path for analytic
extension.
 The amplitude near $Y_c$  for p,p' parallel branesis given by: 

\be
A \simeq V_{p'+1} \int_T^{\infty} {dt \over t}(8\pi^2\alp t)^{-({p'+1 \over 2})}e^{-t\pi Z}\;\;\;\;;\;\;\;\ Z={ Y^2-Y_c^2 \over Y_c^2}. 
\ee
where T is large enough to allow use of the leading term in the integrand.
Hence we have
\be
A \simeq V_{p'+1}(8\pi^2\alp T)^{-(p'+1)/2} \sum_{n=0}^{\infty}\frac{(\pi ZT)^n}{n!}\frac{(-1)^{n+1}}{n-{p'+1 \over 2}}
\ee
for even p (typeIIA theory) and
\be
A \simeq V_{p'+1}(8\pi^2\alp T)^{-(p'+1)/2} \sum_{n=0,\neq (p'+1)/2}^{\infty}
\frac{(\pi ZT)^n}{n!}\frac{(-1)^{n+1}}{n-{p'+1 \over 2}} + 
\frac {(-1)^{{(p'-1)\over 2}}}{{(p'+1) \over2}!}\ln {(\pi TZ)}(\pi TZ)^{(p'+1)/2}
\ee
for odd p (type IIB theory). 
We consider two cases respectively: 

{\bf A)Even P case}
 We can approximate the amplitude near $Y_c$ ($Y >Y_c$) by the first terms of 
the sum (14): 

\be
V\simeq V_{p'+1}(8\pi^2\alp T)^{-(p'+1)/2} [{-2 \over p'+1}+
{4\pi T \over p'-1}({Y-Y_c \over Y_c})+ 
{4\pi^2 T^2 \over 3-p'}({Y-Y_c \over Y_c})^2+ O((Y-Y_c)^3)].
\ee
 Only for the case $p'=0,\; Y_c\neq 0$ the force acting on zero brane at  
$Y=Y_c$ is replusiveand the potential is harmonic and shows a minimum at 
$Y=Y_c (1+{3 \over 2\pi T})$.
 This means that in $D_2$-brane, $D_0$-brane scattering near $Y_c$ $D_0$-brane
feels a harmonic potential, this procedure does not hold for $p'=2$, e.g. 
$D_2$-brane, $D_4$-brane scatteing, in this case force acting on a $D_2$-brane
is always attractive and we do not get a minimum in potential.

{\bf B)odd P case}
 In this case potential is approximated by the first terms of (15) which can be
separated for different p': 
\be
\begin{array}{cc}
p'=-1 (D-instanton) V\simeq -2\pi T({Y-Y_c \over Y_c})+
2\pi^2 T^2 ({Y-Y_c \over Y_c})^2+ln 2\pi T({Y-Y_c \over Y_c})\\
p'=1 \;\;\; V\simeq -1+2\pi T ({Y-Y_c \over Y_c})^2-
2\pi T({Y-Y_c \over Y_c})ln 2\pi T({Y-Y_c \over Y_c})\\
p'=3 \;\;\;\; V\simeq -1/2+2\pi T ({Y-Y_c \over Y_c})+
2\pi^2 T^2({Y-Y_c \over Y_c})^2 ln 2\pi T({Y-Y_c \over Y_c})\\
\end{array}
\ee
and same as (16) for $p'=5$.
 In all of the cases wirtten above force at $Y_c$ is attractive and we never 
reach to a harmonic case as in A) case.
\\
{\bf iii)String D-brane scattering and brane thickness:}

 The same thickness can be viewed from string, $D_p$-brane scattering.

 As it has been discussed in [7,13], a massless string $D_p$-brane scattering
amplitude, is:
\be 
A=\frac{\Gamma(1-2s)\Gamma(-t/2)}{\Gamma(1-2s-t/2)} P \prod_{A=1}^{p+1}\delta(k^A+P^A)
\ee
where P is the related polarization factor and $s,t$ are incoming and transfered
momentum squared respectively.

 This Veneziano type amplitude justifies the stringy spectrum and degeneracy. 
 More over the Fourier transformation of this amplitude with
respect to momentum transfer, which in field theoretical point of view is the
closed string momentum exchanged between brane and massless string, shows the same
thickness we obtained before, $the\ strnig\ scale$. 

 Studying the brane string scattering in the dual channel, we obtain the $brane 
\ spectrum$ to be the same as $open\ string\ spectrum$ (poles of the amlpitude given  
above in t channel). So D-brane excitations are realized by open strings atteched to brane.

 The same results for D-brane spectrum and thickness is obtained in 
decay of the excited branes [8].

 {\bf Discussion}

 We have given evidence to find when the string scale persists in presence of the 
D-branes, and when the halo fomed around them becomes penetrable. Neither  
ordinary string states nor branes even at small velocities have interaction
amplitudes with scale different from the string scale except when $D_0$-branes
probes a $D_0$-brane or a $D_4$-brane.

 Between the two scales , the compton wave length of the brane and the string
scale; the larger one is the one that will be observed. But there are some 
certain D-brane, D-brane cases where the amlpitudes vanishes which at first 
may be thought as a mechanism to supress the string scale. On the other hand
the slightest perturbation in relative orientation of the branes which need 
infinitesimal energy in realistic theories where the brane is wrapped around 
certain cycles of the compact manifolds, reveals the string scale. 

 Moreover in the low energy limit of the theory i.e. SUGRA p-branes also show 
the scales of the same order [3]. This scale is not as hard as one observes 
in the case of D-branes.
 In these low energy theories massive states which are responsible for the singular 
structure of the amlpitudes are not present. 
This is why the scale in extended solutions of the SUGRA
is not accompanied by singularity. As a matter of fact in the points of view 
discussed in this article the singularities can be attributed to large mass 
behaviour of the spectrum of the superstrings. In SUGRA theories these effects
may be due to non-linearities [3].

 The charactristic length obtained is closely related to the Hagedron
temprature . In the same way that approching Hagedron temprature is accompanied by
singularities in string thermodynamic quantities approching distances shorter than
string limit is accompanied by singularity in amplitudes as effective potentials.
Whether this singular nature persists in some more fundamental theory lying
behind string theory is not obvious. If such theory (M or other) is based on 
point-like objects [14] then it may be removed since high energy states 
may loose their
stringy properties and show themselves as statses of point particles.

 On the other hand if these M or other theories use extended objects as strings or
branes as fundamental entites one expects the singular nature to persist.

{\bf Acknowledgement:}

The authors would like to thank C. Bachas and F. Ardalan for fruitful 
discussions.

\end{document}